\documentstyle[preprint,eqsecnum,aps]{revtex}

\begin{document}
\draft
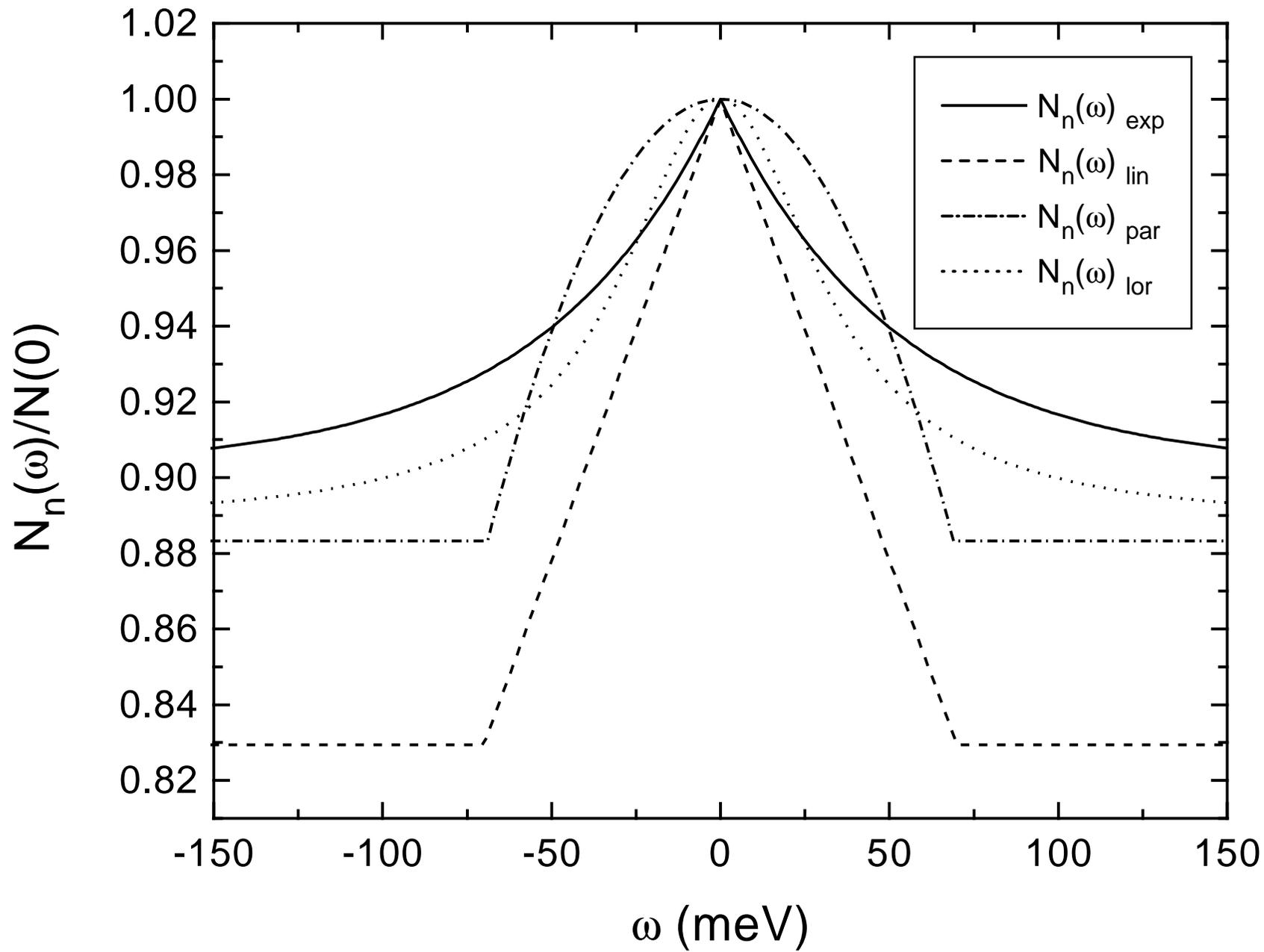
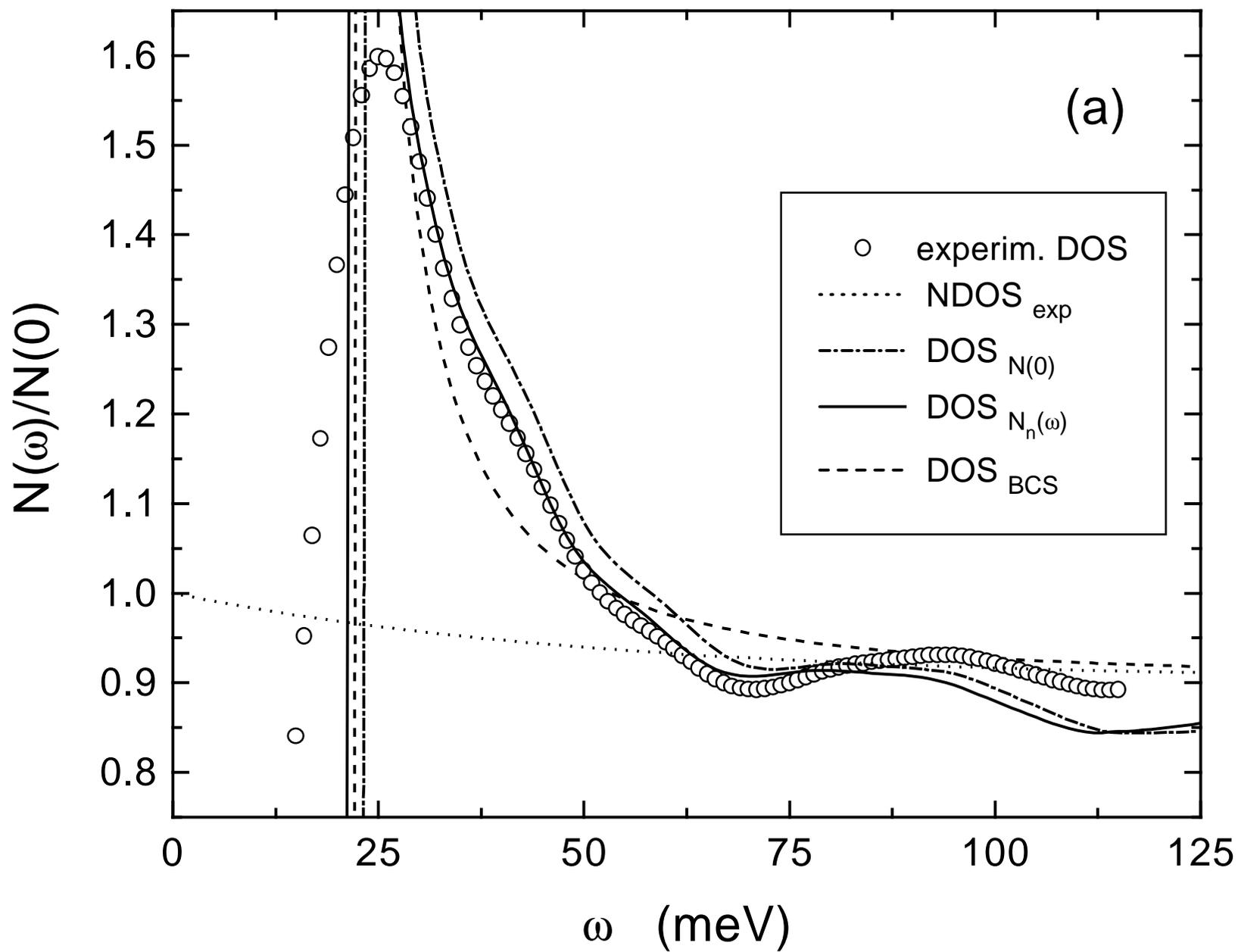
\title{Determination of the tunneling electron-phonon spectral function in high-T$_{%
\text{c}}$ superconductors with energy dependence of the normal density of
states.}
\author{R.S. Gonnelli,$^1$ G.A. Ummarino,$^1$ and V.A. Stepanov$^2$}
\address{$^1INFM-$Dipartimento di Fisica, Politecnico di Torino, 10129 Torino, Italy}
\address{$^2$P.N. Lebedev Physical Institute, Russian Academy of Sciences, Moscow, Russia}
\maketitle

\begin{abstract}
In the present paper we discuss the limits and the correct utilization of
the standard program for the inversion of Eliashberg equations and the
determination of the electron-phonon spectral function and the coulomb
pseudopotential from tunneling measurements in high-$T_c$ superconductors.
In order to compare the calculated density of states with the experimental
one, we introduce the results of the inversion procedure, applied to our
recent tunneling data in Bi$_2$Sr$_2$CaCu$_2$O$_{8+x}$ single-crystal break
junctions with $T_c=93$ K, in a direct program for the solution of
Eliashberg equations. Most of the observed differences between theoretical
and experimental curves at energy greater than gap can be explained by a
smooth energy dependence of the normal density of states that we introduced
in the direct solution of the Eliashberg equations. Finally we show that the
effects of the energy-dependent normal density of states can be simulated by
an efficient electron-phonon spectral function but, also, by a negative,
nonphysical, coulomb pseudopotential.
\end{abstract}


\preprint{HEP/123-qed}

\narrowtext

\section{Introduction}

At the present time it can be regarded as a well-established fact that in
the oxide superconductors, as in conventional ones, and at temperatures
below the critical temperature $T_c$, the charge carriers are bound in pairs
in an energy band whose width is of the order of the gap, but the mechanism
responsible for the attraction between the charges has still not been
identified.

A very important method for obtaining information concerning this aspect is
constituted by tunnel spectroscopy measurements\cite{ref1,ref2,ref3,ref4},
that, through the Eliashberg equations (EE) inversion, allow to determine
the electron-phonon spectral function, $\alpha ^2\left( \omega \right)
F\left( \omega \right) $, of the material. Seven years after the publication
of the first, tentative, $\alpha ^2(\omega )F(\omega )$ curves obtained by
inversion of the gap equations from point-contact tunneling data in Bi$_2$Sr$%
_2$CaCu$_2$O$_{8+x}$ (BSCCO) \cite{ref5,ref6}, a certain skepticism still
persists on the possibility of obtaining a reliable and reproducible $\alpha
^2(\omega )F(\omega )$ from tunneling data in BSCCO. The main reasons for
this skepticism are related to: i) the determination of the low-temperature
normal-state conductance that is a fundamental ingredient for the inversion
of the gap equations, but it is not a measurable quantity in this high-$T_c$
superconductor and ii) the considerable deviations of the $dI/dV$ curves
from the ideal BCS behaviour, observed in this material up to short time
ago. More recently, tunneling experiments using the break-junction approach
have shown that single crystals of very high structural and crystallographic
quality and a very fine technique for the production of stable and
reproducible junctions can lead to BCS-like lifetime-broadened results \cite
{ref3,ref4}. In this case the deviations of the BSCCO break-junction $dI/dV$
characteristics from the ideal Superconductor-Insulator-Superconductor (SIS)
behaviour are relatively small. After an unfolding procedure for extracting
the density of states $N_S$ from the SIS tunneling conductance, it is
possible to observe that most of the deviations of $N_S$ from the BCS
density of states can be explained in all the energy range only by the
presence of a certain amount of lifetime broadening \cite{ref7}. Similar
results have been recently obtained with the point-contact technique by
using highly doped GaAs as counterelectrode \cite{ref2}.

In this work we develop a method for the correct utilization of the best
experimental tunneling data in the inversion of the EE, by introducing
different theoretical energy-dependent normal densities of states (NDOS)
determined by suitable criteria we will discuss in detail in section III. In
this way we have calculated $\alpha ^2(\omega )F(\omega )$, the coulomb
pseudopotential $\mu ^{*}$ and other observable physical quantities that
depend on the first two, for every normal density of states. Then we have
put $\alpha ^2(\omega )F(\omega )$ and $\mu ^{*}$ in a direct program for
the solution of the EE and, for each different NDOS, we calculated the
superconductive quasiparticle density of states (DOS) for comparison with
the experimental data. Some possible explanations for the discrepancies
between theoretical and experimental DOS are given. By using the direct
program and including the effect of an energy-dependent symmetric NDOS of
proper shape, we find that the agreement with the experimental data improves
very much at energies greater than gap. Finally, from the theoretical DOS
calculated by the direct program with an energy-dependent NDOS, and by using
again the standard inversion program we determine the efficient values of $%
\alpha ^2(\omega )F(\omega )$ and $\mu ^{*},$ that simulate the effects of
the NDOS depending on energy. From now on we set $\hbar =c=k_B=1$.

\section{Inversion of the Eliashberg equations}

The experimental data that we use in order to extract the $\alpha ^2(\omega
)F(\omega )$ and to compare with the theoretical DOS have been obtained by
employing the break-junction technique on BSCCO single crystals of very high
crystallographic and chemical quality, as indicated by x-ray diffractograms,
resistivity versus temperature and susceptibility measurements. Before
breaking the crystals, their $R(T)$ characteristics were measured and the
critical temperature was $T_c=93$ K with $\Delta T_c(10-90\%)=2$ K.
Immediately afterwards, at 4.2 K, a finely controlled bending force was
applied to the thin sample holder which the crystal was mounted on. It
produced the break of the very thin crystal (about $1\times 1\times 0.02$ mm$%
^3$) perpendicularly to the $ab$ planes and the creation of the junction. By
varying the bending force it was possible to reproducibly and accurately
control the resistance of the junction and also the $I(V)$ curves were very
stable and reproducible. All the details concerning the experimental data
and the reproducible determination of the density of states and the
Eliashberg function in these crystals can be found in a forthcoming paper 
\cite{ref8}. Among several $I(V)$ tunneling measurements of very high
quality we selected one of the best and applied an unfolding procedure for
extracting the quasiparticle density of states $N_S$ from the SIS tunneling
conductance. Figure 1 shows that the quality of the experimental DOS (open
circles) is largely improved with respect to the measurements of few years
ago and, in particular, sharp peaks at the energy gap $\Delta $ and an
almost constant $N_S$ at $eV\gg \Delta $ are now present. Nevertheless, it
is possible to observe that deviations of $N_S$ from the ''pure'' BCS
density of states $N_{BCS}$ are still present: it is a problem for the
inversion of EE \cite{ref9,ref10,ref11,ref12,ref13,ref14} because very large
oscillations in $N_{rid}=N_S/(N_n\cdot N_{BCS})-1$ (where $N_n$ is the
normal density of states) produce singularity and no convergence at the
increase of the number of iteractions when the standard inversion program is
used \cite{ref13,ref14}. Usually this difficulty in the calculation of the
reduced density of states $N_{rid}$ is overcome by using the
lifetime-broadened expression \cite{ref4,ref5,ref6,ref8} instead of $N_{BCS}$%
: $N_l(\omega )=real\left\{ (\omega -i\Gamma )/\sqrt{(\omega -i\Gamma
)^2-\Delta (0)^2}\right\} $ . The values of the energy gap $\Delta (0)$ and
of the lifetime broadening parameter $\Gamma $ that fit the experimental
density of states of Fig. 1 are $\Delta (0)=23$ meV, $\Gamma =2.9$ meV and,
consequently, $2\Delta (0)/T_c=5.74$. In Fig. 1 the solid line represents
the lifetime broadening fit of the experimental data.

Even if widely used as a first approximation for the $\alpha ^2(\omega
)F(\omega )$ determination, the procedure of using $N_l(\omega )$ instead of 
$N_{BCS}$ appears somehow incoherent. In fact, some parts of the inversion
program contain the ideal reference curve, $N_{BCS}$, but, for calculating $%
N_{rid}$, a different curve $N_l$ is used. For a fully coherent utilization
of the inversion program it is necessary to compare the experimental data to
the ideal BCS behaviour. We also believe that it is fundamental to normalize
the experimental data to an energy-dependent NDOS and then calculate the $%
N_{rid}$ by using the $N_{BCS}$ as reference. In high-$T_c$ superconductors
(HTS) like BSCCO, the normal conductance is unknown at the temperature of
measurement, for example $T_{ex}=4.2$ K, since the upper critical magnetic
field necessary to suppress the superconducting phase is very large and
unrealizable, while, on the other hand, $\Delta T=T_c-T_{ex}$ is too large
to consider NDOS($T_c$)$\simeq $ NDOS($T_{ex}$). For this reason, we will
use four simple arbitrary expressions for the NDOS with three free
parameters each, to be determined by suitable physical conditions. Comparing
the results of this approach to the experimental data of Fig. 1 we will
select the NDOS that, possibly, will be the most plausible one for our
break-junction tunneling experiment.

\section{Energy dependence of the normal density of states}

The HTS have a very complicated crystallographic structure. Ho et al. \cite
{ref15} have given general motivations to the fact that six or more atoms of
the transition elements for unitary cell may produce definite peaks in the
NDOS. There are also many experimental results indicating this hypothesis as
correct\cite{ref8,ref16,ref17}. In particular the results of Mandrus et al.
in BSCCO \cite{ref18} and our recent measurements in Bi$_2$Sr$_2$CuO$_{6+x}$
single crystals suggest an energy dependence of the NDOS which varies with
the direction of the tunneling process ($ab$ plane or $c$ axis). Since the
exact dependence is unknown, we used four different, but symmetric,
analytical forms for the energy dependence of the NDOS: exponential, linear,
parabolic and lorentzian

\begin{equation}  \label{one}
N_n^{\text{exp}}(\omega )=\frac{N(0)}{1+x}\left[ 1+x\cdot \exp (\frac{-\mid
\omega \mid }a)\right]
\end{equation}
\begin{equation}  \label{two}
N_n^{lin}(\omega )=\frac{N(0)}{1+x}\left[ 1+x\cdot \theta (a-\mid \omega
\mid )\cdot (1-\frac{\mid \omega \mid }a)\right]
\end{equation}
\begin{equation}  \label{three}
N_n^{par}(\omega )=\frac{N(0)}{1+x}\left[ 1+x\cdot \theta (a-\mid \omega
\mid )\cdot (1-\frac{\mid \omega \mid ^2}{a^2})\right]
\end{equation}
\begin{equation}  \label{four}
N_n^{lor}(\omega )=\frac{N(0)}{1+x}\left[ 1+\frac{x\cdot a^2}{a^2+\omega ^2}%
)\right]
\end{equation}
where $N(0)$, $a$ and $x$ are free parameters. $N(0)$ is the density of
states at the Fermi level and $x$ controls the behaviour of the DOS at large
energies. In the best tunneling measurements along the $ab$ plane, it is
reasonable to consider $x\geq 0$, i.e. the tunneling conductance is constant
or slightly decreasing at large energies, both for experimental reasons \cite
{ref3,ref4,ref8,ref18} and because the total isotope effect is less than $%
1/2 $ in HTS and this implies $\partial ^2N_n(\omega )/\partial \omega
^2\mid _{\omega =0}<0$ \cite{ref19}. Of course in all the Eqs. 3.1-3.4, $%
N_n(\omega )=N(0)$ when $\omega $ tends to zero.

For determining the values of the parameters we employ the following
conditions:

a) $\lim_{\omega \rightarrow +\infty }N_S(\omega )=\lim_{\omega \rightarrow
+\infty }N_n(\omega )=N_{ex}(+\infty )$

where $N_{ex}(+\infty )$ is the experimental density of states at energy $%
\omega \gg \Delta (0)$;

b) $\int_0^{+\infty }N_{ex}(\omega )d\omega =\int_0^{+\infty }N_n(\omega
)d\omega $

which expresses the law of conservation for the number of states;

c) $\int_{\Delta (0)}^{+\infty }\{N_{ex}(\omega )/[N_n(\omega )\cdot
N_{BCS}(\omega )]-1\}d\omega \simeq 0$

since the oscillations of the experimental ''normalized'' data around the
ideal BCS curve in the phonon energy range should be symmetric \cite
{ref1,ref20}.

The condition b is ideal: in reality the experimental curve shows a small
zero-bias conductance and we have applied this modified condition:

b') $\int_0^{\omega _m}N_{ex}(\omega )d\omega -A_{lc}=\int_0^{\omega
_m}N_n(\omega )d\omega $

where $\omega _{m\text{ }}$is the highest energy of the experimental data
and $A_{lc}=c\cdot \omega _m$ with $0\leq c\leq N_{ex}(0)$ is a small
contribution eventually produced by leakage currents.

In the same way, for the practical estimation of the parameters, the
condition c becomes:

c') $\min \left| \int_{\Delta _m}^{+\infty }\{N_{ex}(\omega )/[N_n(\omega
)\cdot N_{BCS}(\omega )]-1\}d\omega \right| $

where $\Delta _m>\Delta (0)$ is the first point of the experimental data
used in the standard procedure for the inversion of the Eliashberg equations.

In the case of our experimental data shown in Fig. 1, a program for the
numerical calculation of the parameters according to the conditions a, b'
and c' yields :

\begin{center}
$
\begin{tabular}{|c|c|c|c|c|}
\hline
& $N(0)(\Omega ^{-1})$ & $x$ & $a(meV)$ & $A_{lc}(meV\cdot \Omega ^{-1})$ \\ 
\hline
exp & $6.099\cdot 10^{-2}$ & $0.108$ & $52$ & $0.403$ \\ \hline
lin & $6.635\cdot 10^{-2}$ & $0.206$ & $70$ & $0.288$ \\ \hline
par & $6.230\cdot 10^{-2}$ & $0.132$ & $69$ & $0.345$ \\ \hline
lor & $6.200\cdot 10^{-2}$ & $0.127$ & $35$ & $0.368$ \\ \hline
\end{tabular}
$
\end{center}

The different, energy-dependent NDOS curves corresponding to the parameters
of the table are shown in Fig. 2. If we know the $I(V)$ curve for $T\geq T_c$%
, in order to check the quality of the fit for $N_n(\omega )$, we can use
the well-known expression of the tunnel current of a symmetric junction:

$I(V)=e^{-1}G_{NN}\cdot \int_{-\infty }^{+\infty }N_n(\omega )N_n(\omega
+eV)\left[ f(\omega )-f(\omega +eV)\right] d\omega $

where $G_{NN}$ is a constant and $f(\omega )$ is the Fermi function. Since $%
G_{NN}$ is unknown, we can use the quantity $I(V)/[\partial I(V)/\partial V]$
for a comparison of the theoretical results with the experimental data.

Now we can correctly calculate the $N_{rid}=N_{ex}(\omega )/[N_n(\omega
)\cdot N_{BCS}(\omega )]-1$ and, by the standard inversion program \cite
{ref13}, we can determine the spectral function $\alpha ^2(\omega )F(\omega
) $, the coulomb pseudopotential $\mu ^{*}$, the electron-phonon interaction
constant $\lambda =\int_0^{+\infty }[\alpha ^2(\omega )F(\omega )/\omega
]d\omega $, $T_c$, the area $A$ of $\alpha ^2(\omega )F(\omega )$, the
coefficient $b$ of the quadratic part of $\alpha ^2(\omega )F(\omega )$ for $%
\omega \rightarrow 0$ and $\omega _c$, that is the cut-off frequency. The
results of this inversion procedure, applied to the experimental data of
Fig. 1, are summarized in the following table for the different
energy-dependent NDOS curves:

\begin{center}
$
\begin{tabular}{|c|c|c|c|c|c|c|}
\hline
& $\lambda $ & $\mu ^{*}$ & $A(meV)$ & $T_c(K)$ & $b(meV^{-2})$ & $\omega
_c(meV)$ \\ \hline
exp & $3.34$ & $0.93\cdot 10^{-2}$ & $47.21$ & $93.6$ & $3.9\cdot 10^{-3}$ & 
$225$ \\ \hline
lin & $2.86$ & $0.26\cdot 10^{-1}$ & $47.68$ & $99.7$ & $1.6\cdot 10^{-3}$ & 
$225$ \\ \hline
par & $3.03$ & $0.31\cdot 10^{-1}$ & $48.40$ & $98.8$ & $2.7\cdot 10^{-3}$ & 
$225$ \\ \hline
lor & $3.24$ & $0.31\cdot 10^{-1}$ & $47.49$ & $94.6$ & $3.0\cdot 10^{-3}$ & 
$225$ \\ \hline
\end{tabular}
$
\end{center}

In Fig. 3 the spectral functions $\alpha ^2(\omega )F(\omega )$ for the
different NDOS models are compared to the generalized phonon density of
states $G(\omega )$ of BSCCO \cite{ref21}.

By solving the standard Eliashberg equations in direct way we can now check
if the different $\alpha ^2(\omega )F(\omega )$ and $\mu ^{*}$ reproduce the
experimental density of states $N_{ex}(\omega )$. The well-known standard EE
are: 
\begin{equation}  \label{5}
\Delta (\omega )Z(\omega )=\int_0^{\omega _c}P(\omega ^{^{\prime }})\cdot
\left[ K_{+}(\omega ,\omega ^{^{\prime }})-\mu ^{*}\right] d\omega
^{^{\prime }}
\end{equation}
\begin{equation}  \label{6}
\left[ 1-Z(\omega )\right] \cdot \omega =\int_0^\infty N_S(\omega ^{^{\prime
}})\cdot K_{-}(\omega ,\omega ^{^{\prime }})d\omega ^{^{\prime }}
\end{equation}

where 
\[
K_{\pm }(\omega ,\omega ^{^{\prime }})=\int_0^{+\infty }\alpha ^2(\Omega
)F(\Omega )\cdot \left[ \frac 1{\omega ^{^{\prime }}+\omega +\Omega +i\cdot
\delta }\pm \frac 1{\omega ^{^{\prime }}-\omega +\Omega -i\cdot \delta }%
\right] d\Omega 
\]

and $P(\omega )=Real\left( \Delta (\omega )/\sqrt{\omega ^2-\Delta ^2(\omega
)}\right) $is the pair density of states, while $N_S(\omega )=Real\left(
\omega /\sqrt{\omega ^2-\Delta ^2(\omega )}\right) $is the quasiparticle one.

Figures 4 (a), 4 (b), 5 (a) and 5 (b) show a comparison of the densities of
states calculated for $N_n(\omega )=$ cost by solving Eqs. 3.5 and 3.6
(dash-dot lines) with the normalized experimental ones (open circles) and
with the densities of states calculated by the modified EE that take into
account the four different models for $N_n(\omega )$ as it will be discussed
in the following (solid lines). Discrepancies between the theoretical DOS
for $N_n(\omega )=$ cost and the normalized experimental curve are
remarkable for all the different forms of $N_n(\omega )$. They can be due to
the inversion procedure as well as to the form of the Eliashberg equations.

For what concerns the inversion program, it is necessary to remember that:

1) The true normal conductance can be more cumbersome than the theoretical
one and, as a consequence, $N_{rid}$ may be not correct.

2) The standard inversion program was made for ideal BCS-like curves, not
for gapless, broadened ones.

3) The dispersion relations, used in the standard program for calculating $%
Real(\Delta (\omega ))$ and $Im(\Delta (\omega ))$ \cite{ref13,ref14}, are
incorrect if $N_{ex}(0)\neq 0$.

4) In the standard inversion program $\lim_{\omega \rightarrow 0}\alpha
^2(\omega )F(\omega )=\lim_{\omega \rightarrow 0}b\cdot \omega ^2$
(parabolic behaviour) whereas in superconductors with bidimensional
character like BSCCO we can expect a linear behaviour: $\lim_{\omega
\rightarrow 0}\alpha ^2(\omega )F(\omega )=\lim_{\omega \rightarrow 0}a\cdot
\omega $.

5) The energy dependence of the normal density of states is neglected in the
standard inversion program, i.e. $N_n(\omega )=N(0)$.

Concerning the Eliashberg equations it should be born in mind that:

1) The Migdal's theorem is no more valid in HTS since the Debye phonon
frequency is of the order of the Fermi energy \cite{ref22}.

2) The Eliashberg equations in the simple form expressed by Eqs. 3.5 and 3.6
are valid only for homogeneous and isotropic three-dimensional
superconductors whereas BSCCO is bidimensional and very anisotropic.

3) The order parameter is supposed to be in s-wave symmetry.

4) The coulomb pseudopotential is considered constant but it could be energy
dependent \cite{ref23}.

5) In the simple, standard form of the Eliashberg equations $N_n(\omega
)=N(0)$.

In spite of the previous assertions, we assume two fundamental hypotheses:

a) The standard inversion program provides results approximately correct for
what concerns $\alpha ^2(\omega )F(\omega )$ and $\mu ^{*}$.

b) The differences between experimental and theoretical curves at $\omega
>\Delta $ can be explained by putting $N_n(\omega )\neq N(0)$.

The new Eliashberg equations that take into account the energy dependence of
the normal density of states at $T=0$ are \cite
{ref24,ref25,ref26,ref27,ref28}:

\[
\Delta (\omega )Z(\omega )=\int_0^{\omega _c}n_1(\omega ^{^{\prime }})\cdot
\left[ K_{+}(\omega ,\omega ^{^{\prime }})-\mu ^{*}\right] d\omega
^{^{\prime }}, 
\]
\[
\left[ 1-Z(\omega )\right] \cdot \omega =\int_0^{+\infty }n_2(\omega
^{^{\prime }})\cdot K_{-}(\omega ,\omega ^{^{\prime }})d\omega ^{^{\prime
}}, 
\]

\[
\chi (\omega )=\int_0^{+\infty }n_3(\omega ^{^{\prime }})\cdot K_{+}(\omega
,\omega ^{^{\prime }})d\omega ^{^{\prime }}. 
\]

The normal density of states depending on energy $\rho (\omega )=N_n(\omega
)/N(0)$ enters through three functions defined as: 
\[
n_1(\omega )=\left( -\frac 1\pi \right) Im\left( \int_{-\infty }^{+\infty
}\rho (\omega ^{^{\prime }})\cdot \frac{\Delta (\omega )\cdot Z(\omega )}{%
D(\omega ,\omega ^{^{\prime }})}d\omega ^{^{\prime }}\right) , 
\]
\[
n_2(\omega )=\left( -\frac 1\pi \right) Im\left( \int_{-\infty }^{+\infty
}\rho (\omega ^{^{\prime }})\cdot \frac{\omega \cdot Z(\omega )}{D(\omega
,\omega ^{^{\prime }})}d\omega ^{^{\prime }}\right) , 
\]
\[
n_3(\omega )=\left( \frac 1\pi \right) Im\left( \int_{-\infty }^{+\infty
}\rho (\omega ^{^{\prime }})\cdot \frac{\omega ^{^{\prime }}+\chi (\omega )}{%
D(\omega ,\omega ^{^{\prime }})}d\omega ^{^{\prime }}\right) 
\]

where $D(\omega ,\omega ^{^{\prime }})=Z^2(\omega )\cdot \left[ \omega
^2-\Delta ^2(\omega )\right] -\left[ \omega ^{^{\prime }}+\chi (\omega
)\right] ^2$.

Energies are measured from the chemical potential and the real part of $\chi
(\omega )$ represents a shift of the chemical potential due to the
electron-phonon interaction. The energy-dependent electronic density of
states $\rho (\omega )$ modulates the pair density of states $N(0)\cdot
n_1(\omega )$ and the quasiparticle one $N(0)\cdot \left[ n_2(\omega
)-n_3(\omega )\right] $. In this article, only for simplicity, we have
supposed that all the normal densities of states are symmetric: $N_n(\omega
)=N_n(-\omega )$ and this produces $\chi (\omega )=0$.

\section{Results and discussion}

The theoretical $N_S(\omega )/N(0)$ curves for $N_n(\omega )\neq $ cost
(solid lines) are compared to the experimental normalized curves (open
circles) in Figs. 4 (a, b) and 5 (a, b) for exponential, linear, parabolic
and lorentzian cases, respectively. The same figures also show the normal
densities of states $N_n(\omega )$ (dot lines) and $N_{BCS}(\omega )/(1+x)$
(dash lines) for the four above mentioned models. We can see that the
agreement between the theoretical and the experimental normalized DOS is
particularly good for the exponential and the linear cases (Figs. 4 (a) and
4 (b), respectively).

The gap value and the ratio $2\Delta (0)/T_c$ are modified by the energy
dependence of the NDOS as it is shown in the following table:

\begin{center}
$
\begin{tabular}{|c|c|c|}
\hline
& $\Delta (0)(meV)$ & $2\Delta (0)/T_c$ \\ \hline
exp & 21.0 & 5.24 \\ \hline
lin & 20.0 & 4.99 \\ \hline
par & 21.2 & 5.27 \\ \hline
lor & 21.3 & 5.31 \\ \hline
\end{tabular}
$
\end{center}

As it has been predicted in a previous work \cite{ref28}, a negative
curvature of the energy-dependent NDOS produces a reduction of the gap $%
\Delta (0)$.

The best case seems to be the exponential one where $\lambda =3.34$, $\mu
^{*}=0.93\cdot 10^{-2}$, $T_c=93.6$ K. We remind that the experimental value
of $T_c$ is $93$ K. In our opinion, these results are remarkable because $%
\mu ^{*}$ is small but positive, $T_c$ is coincident with the experimental
one and $\lambda $ is not too large: the amorphous low-$T_c$ superconductor
Pb$_{0.5}$Bi$_{0.5}$ has $\lambda =3.00$ and $2\Delta (0)/T_c=5.19$ \cite
{ref10}. Moreover if we calculate the electron-phonon coupling constant by
using the following self-consistent formula \cite{ref15,ref29} for $T=0$ K: $%
\lambda =2\int_0^{+\infty }d\omega ^{\prime }N_n(\omega ^{\prime
})\int_0^{+\infty }d\Omega \frac{\alpha ^2(\Omega )F\left( \Omega \right) }{%
\left( \Omega -\omega ^{\prime }\right) ^2}$ we obtain $\lambda =2.69.$ It
is correct to remember that $T_c$ is calculated by the EE with $N_n(\omega
)=N(0)$, but, since $N_n(\omega )$ has not a definite peak near the Fermi
level, the value of $T_c$ is not significantly affected by the energy
dependence of the NDOS \cite{ref28}. In Fig. 6 the experimental data (open
circles) are compared to the theoretical electron density of states for the
exponential case (solid line) in the full positive energy range. The
agreement of the two curves for $\omega >25$ meV is very good.

Similar results have been obtained in 6 additional break-junction tunneling
curves measured on the same BSCCO single crystals with $T_c=93$ K. The $%
dI/dV $ curves suitable for the analysis previously described were selected
among all the measured data by using some criteria that can be summarized as
follows: i) Presence of sharp and symmetric gap structures; ii) Presence of
symmetric and clear phonon-like structures; iii) Presence of a background
conductance smoothly decreasing with energy at $\omega >25$ meV. In all the
cases the use of the exponential energy dependence of Eq. 3.1 for the normal
density of states gave the best agreement between the theoretical DOS and
the experimental one. The electron-phonon spectral functions $\alpha
^2(\omega )F(\omega )$ obtained by the inversion of the Eliashberg equations
in these 6 additional cases show large similarities to the $\alpha ^2(\omega
)F(\omega )$ of Fig. 3 (solid line) and the calculated average values of $%
\lambda $ and $T_c$ present a small variance of the order of $\pm 11\%$ and $%
\pm 2.7\%$, respectively \cite{ref8}.

An interesting question is whether it is possible to reproduce the tunneling
characteristics for $N_n(\omega )\neq $ cost within the framework of the
usual Eliashberg theory. For trying to give an answer we considered the $%
N_{rid}(\omega )$ calculated by the direct program for $N_n(\omega
)=N_n^{\exp }(\omega )$ and, by using the standard inversion program, found
an effective electron-phonon spectral function $\alpha ^2(\omega )F(\omega
)_{eff}$ and a $\mu _{eff}^{*}$ that simulate the energy dependence of the
normal density of states. The results are shown in Fig. 7 where the
effective electron-phonon spectral function is represented as a dash line
while the original $\alpha ^2(\omega )F(\omega )$ determined from
experimental data (after normalization with $N_n^{\exp }(\omega )$) is shown
as a solid line and the open circles represent the neutron generalized
phonon density of states. We obtained $\lambda _{eff}=3.4$, $\mu
_{eff}^{*}=-0.31\cdot 10^{-1}$, $T_{ceff}=94$ K, $A_{eff}=40.7$ meV, $%
b_{eff}=9\cdot 10^{-3}meV^{-2}$.

Finally, Figure 8 shows the electron-phonon coupling factor $\alpha
^2(\omega )$ as determined from the data shown in Fig. 7. We can observe
that $\alpha ^2(\omega )$ determined from experimental data (solid line) has
a large energy dependence which suggests the presence of a strong
electron-phonon coupling with phonon modes between 12 and 27 meV and between
60 and 80 meV, while the coupling appears rather depressed for phonon modes
between 29 and 56 meV. Similar results have been obtained in previous
break-junction tunneling experiments in BSCCO and can be explained by the
possible directionality mainly along the $ab$ plane of the tunneling process
in our break junctions \cite{ref4,ref8}. Moreover, in Fig. 8 it can be seen
that not to consider the energy dependence of the NDOS (i.e. to use the
standard program for the inversion of the EE on the DOS\ obtained by the
direct solution in presence of the NDOS\ energy dependence) produces an
effective coupling factor $\alpha ^2(\omega )_{eff}$ (dash line) that
overestimates the soft phonon modes and gives rise to a little and negative
nonphysical value of the coulomb pseudopotential.

In conclusion, we have seen that in order to obtain plausible values for $%
\alpha ^2(\omega )F(\omega )$, $\lambda $, $\mu ^{*}$ and $T_c$ it is
necessary to use the standard program for the inversion of EE in a correct
and coherent way. The effect of a nonconstant NDOS on the normalized
tunneling conductance is significant and cannot be accounted for in a
meaningful way within the framework of the standard Eliashberg theory
without obtaining effective parameters $\alpha ^2(\omega )F(\omega )_{eff}$
and $\mu _{eff}^{*}$ that alter the physical interpretation of the problem.
Finally, a strong electron-phonon interaction, restricted to some particular
phonon modes of the $G(\omega )$, and a smooth energy dependence of the NDOS
permit to explain our recent break-junction tunneling data in BSCCO at $%
\omega >\Delta $ in an excellent way. In the next future, it will be
necessary to include in the EE the effect of anisotropy and of the
bidimensional character of Bi-based high-$T_c$ superconductors, as well as,
to calculate the effect of the breakdown of Migdal's theorem on the critical
temperature and on the superconducting gap by using the electron-phonon
spectral functions $\alpha ^2(\omega )F(\omega )$ reproducibly determined in
this and in previous experiments.

\section{Acknowledgements}

Many thanks are due to M. Pescarolo for the help in performing the
break-junction tunneling measurements.

\begin{figure}[tbp]
\caption{Quasiparticle density of states determined from the tunneling
conductance of a single-crystal Bi$_2$Sr$_2$CaCu$_2$O$_{8+x}$ break
junction. }
\label{fig1}
\end{figure}

\begin{figure}[tbp]
\caption{Comparison between the four different energy-dependent normal
densities of states used in this paper and described in Eqs. 3.1 - 3.4. }
\label{fig2}
\end{figure}

\begin{figure}[tbp]
\caption{Spectral functions $\alpha ^2(\omega )F(\omega )$ determined by
using the standard program of inversion of the Eliashberg equations for the
different NDOS models of Fig. 2 compared to the generalized phonon density
of states $G(\omega )$ (open circles).}
\label{fig3}
\end{figure}

\begin{figure}[tbp]
\caption{(a) Comparison of the calculated DOS for $N_n(\omega )=$ cost
(dash-dot line) with the experimental one determined from the data of Fig. 1
(open circles) and with the density of states calculated according to the
NDOS exponential model of Eq. 3.1 (solid line). The BCS density of states
(dash line) and the NDOS (dot line) are also shown for completness; (b) The
same as in (a) but for the linear model of the NDOS.}
\label{fig4}
\end{figure}

\begin{figure}[tbp]
\caption{(a) The same as in Figs. 4 (a) and 4 (b) but for the parabolic
model of the NDOS; (b) The same as in (a) but for the lorentzian model of
the NDOS.}
\label{fig5}
\end{figure}

\begin{figure}[tbp]
\caption{Experimental DOS of BSCCO (open circles) compared to the
theoretical electron density of states for the exponential NDOS model of Eq.
3.1 (solid line).}
\label{fig6}
\end{figure}

\begin{figure}[tbp]
\caption{Comparison of the effective electron-phonon spectral function (dash
line) to the original $\alpha ^2(\omega )F(\omega )$ determined from
experimental data (solid line) and the neutron generalized phonon density of
states (open circles).}
\label{fig7}
\end{figure}

\begin{figure}[tbp]
\caption{Comparison of the original $\alpha ^2(\omega )$ determined from
experimental data in the framework of exponential NDOS model (solid line)
and the effective electron-phonon coupling function $\alpha ^2(\omega
)_{eff} $ (dash line).}
\label{fig8}
\end{figure}

\end{document}